THE EUROPEAN
PHYSICAL JOURNAL C

Regular Article - Experimental Physics

# The measurement of the $H \to \tau\tau$ signal strength in the future $e^+e^-$ Higgs factories

Dan Yu[1], Manqi Ruan[1,a], Vincent Boudry[2], Henri Videau[2], Jean-Claude Brient[2], Zhigang Wu[1], Qun Ouyang[1], Yue Xu[3], Xin Chen[3]

[1] IHEP, Beijing, China
[2] LLR, Ecole Polytechnique, Palaiseau, France
[3] Tsinghua University, Beijing, China

**Abstract** The Circular Electron Positron Collider and the International Linear Collider are two electron-positron Higgs factories. They are designed to operate at a center-of-mass energy of 240 and 250 GeV and accumulate 5.6 and 2 $ab^{-1}$ of integrated luminosity. This paper estimates their performance on the $H \to \tau^+\tau^-$ benchmark measurement. Using the full simulation analysis, the CEPC is expected to measure the signal strength to a relative accuracy of 0.8%. Extrapolating to the ILC setup, we conclude the ILC can reach a relative accuracy of 1.1% or 1.2%, corresponding to two benchmark beam polarization setups. The physics requirement on the mass resolution of the Higgs boson with hadronic decay final states is also discussed, showing that the CEPC baseline design and reconstruction fulfill the accuracy requirement of the $H \to \tau^+\tau^-$ signal strength.

## 1 Introduction

Since the discovery of the Higgs boson in 2012 at the Large Hadron Collider (LHC), the precise measurement of the Higgs boson becomes one of the main goals of the high-energy physics experiments. Compared to the LHC, the $e^+e^-$ Higgs factories offer cleaner collision environment, well known and adjustable initial states, and can determine the absolute value of Higgs boson couplings and total decay width. Many electron-positron Higgs factories are proposed, including the International Linear Collider (ILC) [1], the Circular Electron Positron Collider (CEPC) [2], the Future Circular Collider $e^+e^-$ (FCCee) [3], and the CLIC [4].

The CEPC has a main ring circumference of 100 km and can be operated as a Z factory ($\sqrt{s} = 91.2\ GeV$) and a Higgs factory ($\sqrt{s} = 240\ GeV$). It can also perform a W threshold scan at $\sqrt{s} = 160\ GeV$ and determine the mass and width of the W boson accurately. With a nominal integrated luminosity of 5.6 ab$^{-1}$, the CEPC is expected to produce one million of Higgs bosons [5]. Benefitting from its very clean collision environment, its detector system can record almost all the Higgs events. This pure, large-statistic Higgs sample provides crucial information on top of the Higgs program at the HL-LHC, and can boost the precision of Higgs boson property measurements by up to one order of magnitude [5]. More details can be found in the CEPC Conceptual Design Report released in 2018 [2].

Another $e^+e^-$ Higgs factory, the ILC, has been intensively studied in the past 20 years; the ILC TDR [1] was published in 2013. Compared to circular colliders, linear colliders as the ILC have greater potential for achieving much higher centre-of-mass energies. In the proposed staging scenario [6], the ILC starts collisions at a centre-of-mass energy of 250 GeV, serving as a Higgs factory with a nominal luminosity of 2 $ab^{-1}$. The ILC is foreseen to upgrade its centre-of-mass energy to 380 GeV, 500 GeV, and eventually 1 TeV. These high-energy collisions open the $t\bar{t}$, the $t\bar{t}H$, the $\nu\nu HH$ and the $ZHH$ channels, and also improves the Higgs width measurements significantly. Another notable advantage of the ILC is the capability of beam polarization. Since the left and right-handed fermions have different quantum numbers in the electroweak interaction, the beam polarization provides a degree of freedom to the initial state of the collision, significantly enhancing the physics performance, e.g., for the $sin^2(\theta_w)$ measurements. In the ILC TDR, there are two official settings of the ILC beam polarization denoted as the left and right-hand polarization settings, P(−0.8, 0.3) and P(0.8, −0.3), where the first/second number represents the electron/positron polarization, and the minus sign refers to the left-hand polarization. In terms of the Higgs property measurement, the polarization could also enhance the signal yields, and suppress the SM background. Table 1 indicates the inclusive cross-section, the nominal luminosity, and the expected total Higgs events at the CEPC and the ILC.

[a] e-mail: ruanmq@ihep.ac.cn

**Table 1** The comparison of Higgs signal for the CEPC and the ILC

|  | CEPC | ILC (0.8, −0.3) | ILC (−0.8, 0.3) |
|---|---|---|---|
| Polarization | – | (0.8, −0.3) | (−0.8, 0.3) |
| Luminosity | 5 $ab^{-1}$ | 2 $ab^{-1}$ | 2 $ab^{-1}$ |
| Higgs Cross section | 203.66 | 311.99 | 205.24 |
| Higgs Yield | 1,018,300 | 623,980 | 410,480 |

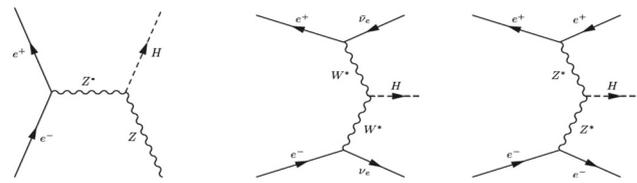

**Fig. 1** Feynman diagram of Higgs production at the CEPC

**Table 2** The cross section ($fb$) of Higgs signal and background for the CEPC and the ILC

| Process | CEPC | ILC (0.8, −0.3) | ILC (−0.8, 0.3) |
|---|---|---|---|
| $eeH$ | 7.04 | 10.69 | 7.14 |
| $\mu\mu H$ | 6.77 | 10.41 | 7.02 |
| $\nu\nu H$ | 46.29 | 77.53 | 42.59 |
| $qqH$ | 136.81 | 202.41 | 141.39 |
| $2f$ | 79,681 | 116,223 | 81,198 |
| Single $Z$ | 4733 | 1817 | 1439 |
| Single $W$ | 5144 | 7865 | 594 |
| $WW$ | 15,483 | 20,614 | 1422 |
| $ZZ$ | 1033 | 1794 | 933 |
| Mixed | 3899 | 8740 | 298 |

Due to the large $\tau$ mass (heaviest SM lepton), a significant fraction of the SM Higgs boson decays into di-$\tau$ final states, making the $H \to \tau\tau$ channel a sensitive probe to the new physics. In [7] this paper, the expected accuracy of the $H \to \tau\tau$ signal strength (defined as anticipated number of events times decay rates relative to Standard Model (SM) expectations) measurement is analyzed using the official CEPC software and samples. Two different analysis methods are developed, corresponding to the signal with or without jets in the final state. For the events without jets ($llH$ and $\nu\nu H$), the signal identification strongly relies on the event multiplicity information. For the events with jets ($qqH$), a dedicated tau-finding algorithm *TAURUS* has been developed (see in Appendix A ), and the signal is identified using the measurements from both the di-tau system and the recoiling di-jet system. The results of all those channels are combined, leading to a relative accuracy of 0.8% of the objective signal strength at the CEPC. Extrapolated to the ILC, this analysis yields an accuracy of 1.1%/1.2% for the nominal left/right-hand polarization setting.

This paper is organized into five sections. Section 2 describes the detector model, the software, and the samples used in this study. Section 3 presents the CEPC analyses at different channels. Section. 4 provides the combination of different channels and extrapolates the result to the ILC setups. The conclusion is summarized in Sect. 5.

## 2 Samples and software chain

The SM Higgs bosons are mainly generated via the Higgsstrahlung and the vector boson fusion processes at the $e^+e^-$ colliders, see Fig. 1. At the designed center-of-mass operation for Higgs production, for both the CEPC and the ILC, the inclusive SM background within the detector fiducial volume is roughly 2–3 orders of magnitudes higher than that of the Higgs signal. In our analysis, the backgrounds are characterized according to the number of final state fermions at the Parton level. At 240 or 250 GeV centre-of-mass energy, the leading SM backgrounds are the 2-fermion and 4-fermion backgrounds. The 2-fermion backgrounds are the $qq$, Bhabha, $\mu\mu$ and $\tau\tau$ processes; the 4-fermion backgrounds include the $ZZ$, the $WW$, the single $W$, the single $Z$, and the interfering processes. The latter is denoted as $ZZorWW$ and $ZorW$ process since the final state fermion combination allows multiple intermediate states and their interferences. The cross sections for signal and backgrounds are summarized in Table 2.

The detector model used in the simulation is the CEPC baseline detector [2], a Particle Flow Oriented detector. It is composed of a low-material tracking system, a high granularity calorimeter system, and a 3-Tesla large radius solenoid that hosts both ECAL and HCAL inside. A baseline CEPC simulation-reconstruction software has been established. It uses the Whizard as the generator [8], the MokkaPlus [9] for the full detector simulation, the Clupatra [10] for tracking, and the Arbor [11] for the PFA reconstruction.

Using the CEPC baseline geometry, an official Monte-Carlo sample production is performed, corresponding to the nominal setting of the CEPC Higgs runs. The samples are scaled according to the luminosity for this massive production. For the Higgs processes with small cross-section, typically under 20 $fb$, the sample is simulated to a minimal statistic of 100 k. For leading 2-fermion standard model background, the production only simulates a fraction (20%) of the expected statistics, to save the computing resource. For all 4-fermion backgrounds, the samples are generated with full statistics.

**Table 3** The statistics of Higgs decaying to $\tau$ in different channels (5.6 ab−1)

| | $\mu\mu H$ | $eeH$ | $\nu\nu H$ | $qqH$ |
|---|---|---|---|---|
| | 2388 | 2483 | 16331 | 48266 |

## 3 Signal strength measurement at the CEPC

We classify the $H \to \tau^+\tau^-$ signal into two categories according to their final states: without jets ($\mu\mu H$, $eeH$ and $\nu\nu H$) and with jets ($qqH$). The expected accuracy of the signal strength in each channel is analyzed independently. The $\tau\tau H$ channel are not covered in this paper. The statistics of the $H \to \tau\tau$ signal at different channels are listed in Table 3.

For the $qqH$ channel, *TAURUS*, a dedicated $\tau$ finding algorithm, has been developed and optimized for this analysis. *TAURUS* identifies all the $\tau$ candidate in an event, from which the Higgs decay products are identified using the pair with leading energy. The remaining particles are recognized as the di-jet system, whose invariant and recoil masses are also used to distinguish the signal from the background.

No specific $\tau$ finding algorithm is used for the events without jets. Instead, these signals are identified firstly using the multiplicity of the charged particles and the photons. For the $\mu\mu H$ and the $eeH$, the prompt leptons are identified using their invariant and recoil mass information, which significantly reduces the SM background.

Roughly 40% of SM $\tau$s decay into a single charged particle and neutrino(s). To ensure a high signal efficiency, the isolated charged particles are intentionally identified as $\tau$ candidates in all those analyses. On the other hand, the CEPC baseline detector is equipped with a high precision vertex system. Because the $\tau$ lepton has a $c\tau$ of 89 μm, the leading tracks decayed from the $\tau$ candidates have a significant impact parameter. Therefore, the track impact parameter is used to distinguish the 1-prong decayed $\tau$ lepton from the prompt isolated tracks.

The analyses of different sub-channels are discussed in detail in the following sections.

### 3.1 $\mu\mu H$ channel

The analysis of the $\mu\mu H$ channel is presented in this section. The cut chain is shown in Table 4.

The prompt $\mu$ pair is a critical signature of the signal. The baseline CEPC detector provides a high-efficiency and high purity identification of the leptons. Requesting at least a pair of $\mu$ preserves 97% of the signal and reduces the entire SM background by 40 times. The prompt di-$\mu$ system is identified as the combination with the closest mass to the 91.2 GeV. The backgrounds are further reduced by applying a constraint on the invariant and recoil mass of the di-$\mu$ system. These requirements suppress the SM background by another 50 times, and the remaining backgrounds are dominated by $2f$ events. Using the selection condition defined in Table 4 (3 first lines), the signal efficiency is 88.5%, and the background rejection rate is 99.95%.

The remaining particles are identified as the di-$\tau$ system. The leading charged particle is identified, and all the particles within one radius to the leading particle are identified as one $\tau$ candidate, while the remaining particles are identified as the other $\tau$ candidate. Since the $\tau$ lepton decays into a small number of charged particles and photons ($\pi^0$ decayed or charged particle radiated), the charged particle and photon multiplicity of each $\tau$ candidate is required to be smaller than 6 and 7, respectively. At this step, the 2-fermion backgrounds, including $\mu\mu$ and $\tau\tau$ (mainly 1-prong after the $\mu$ selection), are suppressed by requiring the existence of the $\tau$ candidates after the $\mu$-pair elimination.

Several variables are extracted from the di-$\tau$ system and combined using the TMVA tool, [12] to suppress the background further. These variables include

- the angle between the leading tracks and the remotest track in each candidate;
- the angle between the leading tracks and the remotest photon in each candidate;
- the angle between the leading photon and the remotest photon in each candidate.

The impact parameter of the leading track in the $\tau$ candidates is also used in the TMVA. By looking at the impact parameters of the tracks, those stemming from $\tau$ decays are further away from the vertex than the others. The impact parameters, along with the transverse and longitudinal directions (D0 and Z0[1]) of the leading track from each $\tau$ candidate, can be extracted.

The BDT distributions for the signal and the SM backgrounds events are shown in Fig. 2. After an optimized cut at BTD value of 0.78, the background is reduced 3 times, at the cost of losing 5% of the signal statistic. See the 7th line of Table 4. After the TMVA event selection in Table 4, the remaining backgrounds are the single $Z$ and $ZZ$ events.

The invariant mass of the $\tau$ pair is calculated using the collinear approximation (assuming the momentum of neutrino or neutrinos is proportional to the $\tau$'s). A small fraction (3%) of the signal events might take a negative value of the $M_{\tau\tau}^{col}$, due to the anomalous reconstructed visible or missing momentum. (non-validity of assumptions will cause the mass spreads but not negative). Meanwhile, in many backgrounds where no $\tau$ exists, the approximation is not valid and also leads to negative $M_{\tau\tau}^{col}$. For simplicity, those events are excluded from the event selection. The distribution of the invariant mass of the $\tau$ pair is shown in Fig. 3 for signal and SM inclusive background with a fit using PDF shape.

---
[1] The impact parameter D0 is the signed distance from the origin to the point of closest approach in the $r - \phi (x - y)$ plane. The impact parameter Z0 is the Z position of the perigee.

**Table 4** $\mu\mu H$ cut flow

| Total generated | $\mu\mu H\tau\tau$ | $2f$ | $SW$ | $SZ$ | $WW$ | $ZZ$ | mixed | $ZH$ | Total Bkg | $\sqrt{S+B}/S(\%)$ |
|---|---|---|---|---|---|---|---|---|---|---|
|  | 2388 | 801,152,078 | 19,517,399 | 9,072,946 | 50,826,211 | 6,389,424 | 21,839,941 | 1,102,582 | 909,900,581 | 1263.17 |
| $N_{\mu^+} > 1$, $N_{\mu^-} > 1$ | 2341 | 22,894,549 | 37,923 | 720,547 | 1,335,231 | 831,861 | 1,251,657 | 567,636 | 27,639,404 | 233.56 |
| $115 GeV < M_{recoil} < 160 GeV$ | 2186 | 864,849 | 154 | 155,502 | 396,485 | 112,837 | 164,225 | 3114 | 1,697,166 | 61.75 |
| $60 GeV < M_{invariant} < 105 GeV$ | 2118 | 662,042 | 0 | 31,145 | 111,376 | 56,642 | 99,874 | 987 | 962,066 | 48.08 |
| $E_{Le} < 65 GeV$ | 2101 | 658,199 | 0 | 17,760 | 111,340 | 56,516 | 99,822 | 957 | 944,594 | 48.02 |
| $N_{Trk}(A/B) < 6$ & $N_{Ph}(A/B) < 7$ | 1977 | 78 | 0 | 996 | 2576 | 8019 | 29 | 105 | 11,803 | 6.16 |
| BDT>0.78 | 1891 | 0 | 0 | 264 | 231 | 3682 | 9 | 39 | 4225 | 4.26 |
| $M^{col}_{\tau\tau} > 0$ | 1853 | 0 | 0 | 259 | 88 | 3099 | 9 | 35 | 3490 | 4.07 |
| $\tau\tau$ invariant mass fit result |  |  |  |  |  |  |  |  |  | 2.75 |

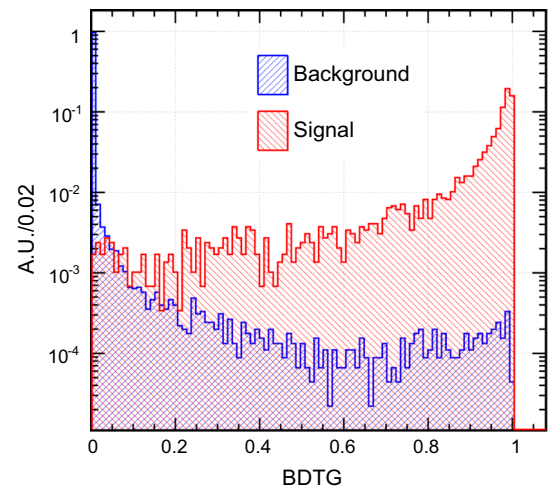

**Fig. 2** The BDT values for the signal and the SM backgrounds events

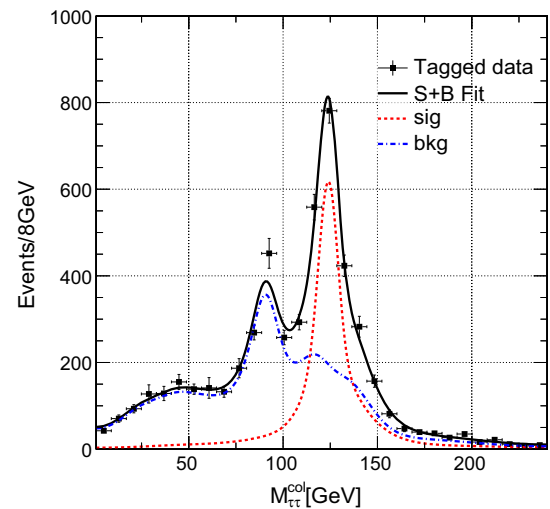

**Fig. 3** Fit of the invariant mass of the di-$\tau$ with SM background included

The number of expected signal event is then fitted from the $\tau\tau$ collinear mass distribution, which has an relative accuracy of 2.75% ($\Delta(\sigma \times BR)/(\sigma \times BR) = 2.75\%$, where the $\delta(S)$ is the fitted signal event number error.)

### 3.2 $eeH$ channel

The analysis of the $eeH$ channel is similar to that of the $\mu\mu H$ channel. Because of the Z fusion events, the signal statistic of the $eeH$ channel is 4% more than the $\mu\mu H$ channel. The electron identification efficiency performance is slightly worse than the muon identification because the electrons have a much significant bremsstrahlung effect. The $eeH$ channel analysis also has much severe single $W$ backgrounds. The cut chain for the $eeH$ channel is shown in Table 5. After optimizing the parameters in the cut chain, 57% of the signal (1422) events survive, and the entire SM background is reduced to a 12k statistics, leading to a relative uncertainty of 8.3%, about 50% worse than that of the $\mu\mu H$ channel analysis.

Similar to the $\mu\mu H$ channel analysis, a fit is performed on the $\tau\tau$ invariant mass. The signal strength of H→ $\tau\tau$ measured from the $eeH$ channel reaches 5.07%.

### 3.3 $\nu\nu H$ channel

The $\nu\nu H$ signal has no jet in its decay final state and has a statistic 6 times larger than the $\mu\mu H$ signal. However, the $\nu\nu H$ has no prompt lepton pairs with clean invariant and recoil mass signature, and the background for the $\nu\nu H$ channel is much larger. All the SM processes with a pair of $\tau$ candidates in the final states and significant missing mass turn to be the irreducible backgrounds, including $2f$ (especially $\tau\tau$), $WW/ZZ$/single $Z$ and the interference events with $\nu\nu\tau\tau$ final states. Besides, induced by the $\tau$ finding strategy, the processes such as single $W$ with the final states $e\nu\tau\nu$ also has a significant chance to be misidentified as the signal.

For the $\nu\nu H$ channel, the parameters for event selection are the missing mass (by recoiling the visible 4-momentum), the transverse momentum, and the total mass. The steps for di-$\tau$ tagging are similar to the $\mu\mu H$ channel, except for that there is no need to veto the lepton pair. The cut chain of $\nu\nu H$ channel is shown in Table 6. After the $\tau$ candidates found, the angles between the $\tau$s are applied to reduce the $2f$ backgrounds. After the cut chain, 55% of the signal events survives, but the background is two orders of magnitude larger than in the $\mu\mu H$ channel, this makes the signal strength accuracy of the $\nu\nu H$ channel two times worse than in $\mu\mu H$ channel.

In this channel, the collinear approximation cannot be used, so only the statistic result of 7.9% from the cut chain is used as the accuracy.

### 3.4 $qqH$ channel

The $qqH$ channel is critical for the $H \to \tau\tau$ signal strength measurement since 70% of the Higgs events at the CEPC are generated via this channel. The cut chain for this analysis is summarized in Table 7. It includes four steps, corresponding to the information of the general event description, the di-$\tau$ system, the di-jet system, and the vertex system correspondingly.

The first step uses the information of the charged particle multiplicity, the total visible energy, and the leading lepton energy. The multiplicity of the charged particles is required to be larger than 10. This requirement eliminates the full leptonic SM background efficiently. According to the visible

**Table 5** $eeH$ cut flow

| | $eeH\tau\tau$ | $2f$ | $SW$ | $SZ$ | $WW$ | $ZZ$ | Mixed | ZH | Total Bkg | $\sqrt{S+B}/S$ (%) |
|---|---|---|---|---|---|---|---|---|---|---|
| Total generated | 2483 | 801,152,078 | 19,517,399 | 9,072,946 | 50,826,211 | 6,389,424 | 21,839,941 | 1,101,070 | 909,899,069 | 1214.84 |
| $N_{e^+} > 1, N_{e^-} > 1$ | 2361 | 252,785,838 | 10,920,426 | 2,069,390 | 4,793,593 | 226,473 | 2,562,603 | 519,007 | 273,877,330 | 798.3 |
| $110 GeV < M_{recoil} < 180 GeV$ | 2051 | 8,931,425 | 3,925,254 | 683,298 | 193,596 | 16,732 | 59,181 | 2232 | 13,811,718 | 206.13 |
| $40 GeV < M_{invariant} < 180 GeV$ | 1943 | 3,046,082 | 643,288 | 337,928 | 51,155 | 4422 | 195,532 | 859 | 4,279,266 | 121.35 |
| $N_{Trk}(A/B) < 6 \& N_{Ph}(A/B) < 7$ | 1828 | 4729 | 22,737 | 22,453 | 4410 | 474 | 651 | 107 | 136,561 | 23.26 |
| BDT > 0.78 | 1619 | 2533 | 3150 | 6315 | 225 | 175 | 271 | 39 | 12,708 | 8.35 |
| $M_{\tau\tau}^{col} > 0$ | 1543 | 2145 | 1894 | 5880 | 27 | 144 | 180 | 27 | 10297 | 7.97 |
| $\tau\tau$ invariant mass fit result | | | | | | | | | | 5.07 |

**Table 6** nnH cut flow

| Total generated | $\nu\nu H\tau\tau$ | 2f | SW | SZ | WW | ZZ | Mixed | ZH | Total Bkg | $\sqrt{S+B}/S$ (%) |
|---|---|---|---|---|---|---|---|---|---|---|
|  | 16,331 | 801,152,078 | 19,517,399 | 9,072,946 | 50,826,211 | 6,389,424 | 21,839,941 | 841,846 | 909,639,845 | 184.68 |
| $110 GeV < M_{missing} < 225 GeV$ | 15,709 | 48,775,523 | 2,140,070 | 1,600,084 | 2,357,138 | 651,545 | 1,795,752 | 9307 | 57,329,419 | 48.21 |
| $M_{total} > 20 GeV$ | 14874 | 12,307,462 | 1,879,196 | 1,155,787 | 1,080,687 | 565,093 | 1,525,347 | 6811 | 1,8520,383 | 28.94 |
| $10 GeV < p_T < 80 GeV$ | 13010 | 8,728,105 | 1,393,911 | 867,874 | 612,205 | 357,230 | 1,250,288 | 5964 | 13,215,577 | 27.95 |
| $E_{Le} < 45 GeV, E_{L\mu} < 65 GeV$ | 11,898 | 7,003,289 | 691,168 | 750,255 | 222,075 | 343,815 | 709,917 | 5681 | 9726,200 | 26.22 |
| $N_{Trk}(A/B) < 6$ & $N_{Ph}(A/B) < 7$ | 10363 | 2,858,665 | 510,018 | 145,838 | 135,057 | 69,682 | 608,076 | 1398 | 4,328,734 | 20.10 |
| BDT > 0.78 | 9960 | 862,244 | 270,754 | 59,187 | 51,522 | 32,776 | 354,743 | 405 | 1,631,631 | 12.86 |
| $2 < \theta_{\tau\tau} < 3$ | 9551 | 439,075 | 107,021 | 35,780 | 40,216 | 17,950 | 141,658 | 381 | 782,081 | 9.3 |
| $2 < \delta\phi_{\tau\tau} < 3$ | 9007 | 206,717 | 94,070 | 29,353 | 39,593 | 14,229 | 114,486 | 357 | 498,805 | 7.9 |

**Table 7** Cut Flow of MC sample for $qqH \to \tau\tau$ selection on signal and inclusive SM backgrounds, $E_{Le}/E_{L\mu}$ represents the energy of the leading electron or muon, $M_{\tau\tau}^{col}$ is the $\tau\tau$ mass calculated with collinear approximation, Pull1 and Pull2 are the pulls of the leading $\tau$ pairs

|  | $qqH\tau\tau$ | 2f | SW | SZ | WW | ZZ | Mixed | ZH | Total Bkg | $\sqrt{S+B}/S$ (%) |
|---|---|---|---|---|---|---|---|---|---|---|
| Total Statistic | 48,266 | 801,152,078 | 19,517,399 | 9,072,946 | 50,826,211 | 6,389,424 | 21,839,941 | 374,357 | 909,679,268 | 62.43 |
| NCh>10 | 47,347 | 272,992,986 | 13,765,307 | 1,969,972 | 47,052,263 | 5,756,249 | 18,020,636 | 331,843 | 359,889,260 | 40.07 |
| $110 GeV < E_{tot} < 235 GeV$ | 46,183 | 173,589,861 | 13,159,096 | 942,644 | 31,297,172 | 3,239,464 | 5,154,115 | 264,535 | 227,646,887 | 32.67 |
| $E_{Le} < 45 GeV, E_{L\mu} < 65 GeV$ | 44,093 | 169,589,868 | 3,413,790 | 707,027 | 22,428,227 | 2,911,836 | 4,985,026 | 237,240 | 204,273,014 | 32.41 |
| $N_{\tau+} > 0, N_{\tau-} > 0$ | 24,214 | 401,147 | 212,183 | 13,999 | 1,129,502 | 171,380 | 193,055 | 16,821 | 2,138,087 | 6.55 |
| $90 GeV < M_{\tau\tau}^{col} < 160 GeV$ | 17,176 | 9717 | 21,483 | 1689 | 135,538 | 62,721 | 7722 | 5305 | 244,175 | 2.97 |
| $70 GeV < M_{qq} < 110 GeV$ | 16,257 | 1596 | 4119 | 1012 | 26,823 | 52,307 | 1818 | 717 | 88,392 | 1.98 |
| $M_{qq}^{rec}(GeV) > 100 GeV$ | 16,211 | 0 | 1463 | 637 | 11,071 | 13,814 | 1265 | 647 | 28,897 | 1.31 |
| 2-D impact parameter fit result |  |  |  |  |  |  |  |  |  | 0.93 |

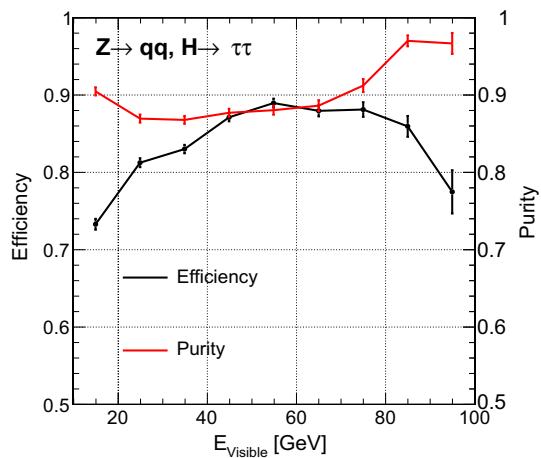

**Fig. 4** $\tau$ finding purity and efficiency at the $qqH \to \tau\tau$ channel

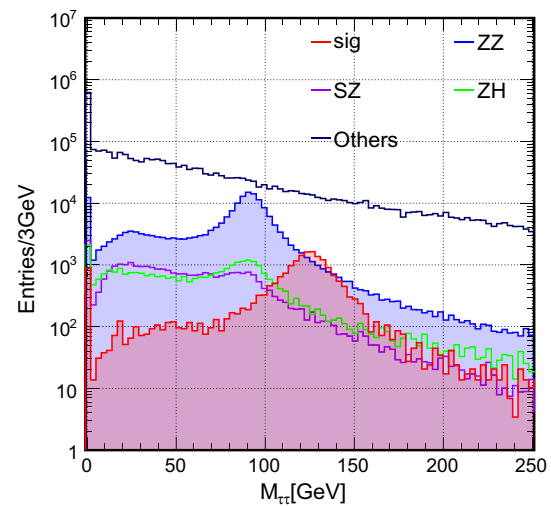

**Fig. 5** Distribution of the invariant mass of the di-$\tau$, $M_{\tau^+\tau^-}$ for $Z \to qq, H \to \tau\tau$, and each backgrounds at $\sqrt{s} = 240$ GeV

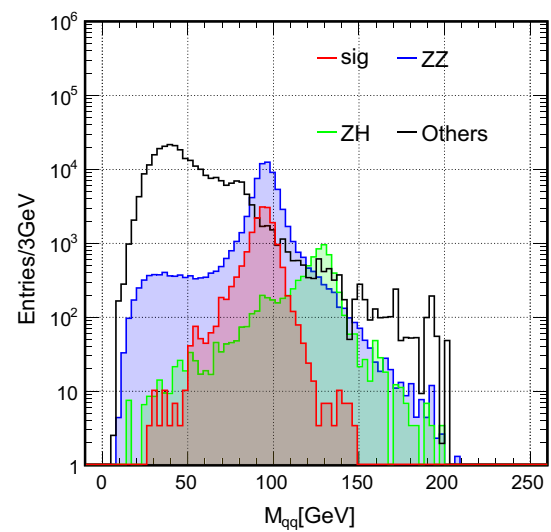

**Fig. 6** Distribution of the invariant mass of the $qq$, $M_{qq}$ for $Z \to qq, H \to \tau\tau$ and each backgrounds at $\sqrt{s} = 240$ GeV after the previous cuts

energy distribution at the signal, the total visible energy is limited to (110, 235) GeV. The upper range limit of 235GeV efficiently reduces the fully visible hadronic events, such as the $ee \to jj (j = uds)$ and the $WW/ZZ \to 4q$. However the statistics of $WW \to \tau q$ and $WW \to \mu q$ is rather large and only less than a half of them can be reduced by lower range limit. The single $W$ and single $Z$ boson backgrounds have energetic final state leptons and are efficiently vetoed reduced by an up limit on the leading lepton energy.

A $\tau$ finding algorithm, TAURUS, is used to identify the $\tau$ candidates. Its parameters are optimized for the signal strength analysis at the $qqH$ channel. An overall $\tau$ finding efficiency/purity of 80%/90% is achieved at the $qqH, H \to \tau\tau$ signal, see Fig. 4. More details can be found in Appendix A. The leading $\tau$ candidates of both charges, if exist, are identified as the decay products of the Higgs boson. The invariant mass of this pair is calculated using the collinear approximation (assuming the momentum of neutrino or neutrinos is proportional to the tau's), as shown in Fig. 5. The second step requires a pair of $\tau$ candidates, and its invariant mass is limited to (90, 160) GeV.

The first step is a gentle selection that preserves 90% of the signal and reduces the SM background by almost 4 times. The second step reduces the remaining background by 3 orders of magnitudes at the cost of losing 60% of the signal events. The second step, especially the $\tau$ finding performance, is critical for this measurement.

After the first two steps, the main backgrounds include the backgrounds with the same final states at the parton level: $ZZ \to qq\tau\tau$ and $ZH(Z \to \tau\tau, H \to qq)$. The signal here is $ZH(Z \to qq, H \to \tau\tau)$, this makes $ZH(Z \to \tau\tau, H \to qq)$ a background. A few $WW$, single $W$ backgrounds survive after the previous steps, where one of the $\tau$ candidates might be generated from the misidentification of TAURUS. These backgrounds can be significantly reduced by using the information of the remaining final state particles, which is defined as the di-jet system. The $ZH$ and $WW$ background can be reduced using the invariant mass of the di-jet system, since the signal peaks at the Z boson mass, while the $ZH$ background peaks at the Higgs mass and the $WW$ background has a flat distribution, see Fig. 6. After the restriction on di-jet invariant mass (line 7 of Table 7), the $ZZ$ background became a dominant one since its di-jet invariant mass also peaks at Z boson mass. However, the recoil-mass of the di-jet system can clearly separate the signal from the $ZZ$ background, see Fig. 7 (line 8 of Table 7).

The PFA oriented detector design and reconstruction provide an accurate reconstruction of the di-jet system. Using the invariant and recoil mass of the di-jet system, the backgrounds are suppressed by one order of magnitude, and the cost of lost 5% of the remaining signal, and the accuracy improves more than a factor of 2, see line 7 and line 8 of Table 7. After these event selections, the dominant backgrounds are $WW$ and $ZZ$, especially their semi-leptonic decay.

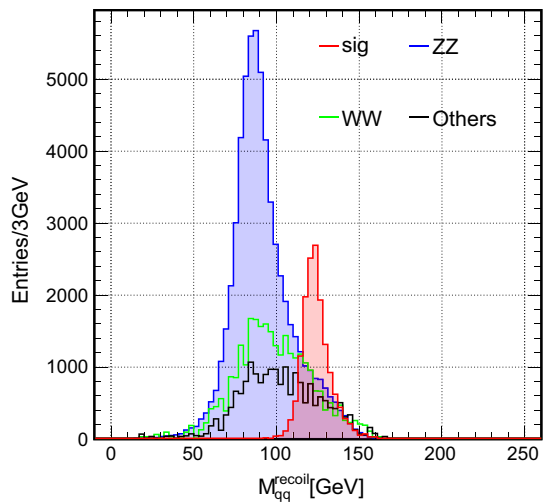

**Fig. 7** Distribution of the recoil mass of the $qq$, $M_{qq}^{recoil}$ for $Z \to qq$, $H \to \tau\tau$ and each background at $\sqrt{s} = 240$ GeV after the previous cuts

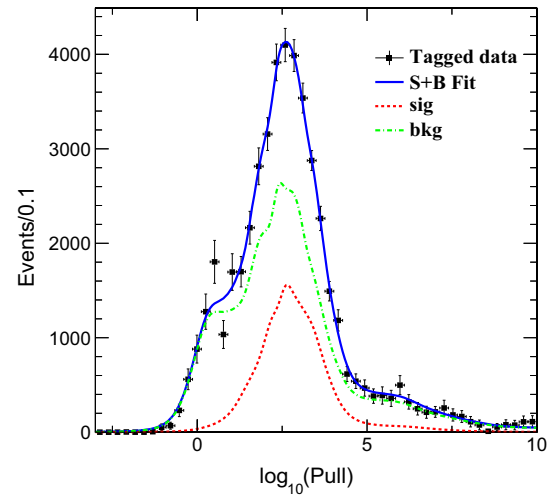

**Fig. 8** Fit of the sum of D0$^2$ and Z0$^2$ of the leading tracks of two cones with SM background included

**Table 8** Combined signal strength accuracy

|  | $\delta(\sigma \times BR)/(\sigma \times BR)$ (%) |
| --- | --- |
| $\mu\mu H$ | 2.8 |
| $eeH$ | 5.1 |
| $\nu\nu H$ | 7.9 |
| $qqH$ | 0.9 |
| Combined | 0.8 |

For each track, the pull is defined as: $D_0^2/\sigma_{D_0}^2 + Z_0^2/\sigma_{Z_0}^2$, where $\sigma_{D0/Z0}$ is the uncertainty of D0 and Z0. The pull parameter of the event is then extracted as the sum of the two $\tau$ candidates decayed from Higgs, which distinguishes the $\tau$ candidates from the prompt tracks, see Fig. 8. In this case, the signal and the background is mixed with each other but their shape can be defined in the later data driven, this makes it possible to use fit method to get a more precise statistic of signal and backgrounds. The signal strength accuracy of the $H \to \tau\tau$) is extracted to be 0.93%.

## 4 Combination of results and extrapolation

To first order, the measurements of different channels ($eeH$, $\mu\mu H$, $\nu\nu H$ and $qqH$) are independent. The signal strength accuracy of Higgs decaying to $\tau\tau$ can be summarized as in Table 8. A total accuracy of 0.8% is achieved with 5.6 $ab^{-1}$, the nominal integrated luminosity of the CEPC Higgs runs. In the $eeH$ and the $\nu\nu H$ channel, it is shown that the accuracy is significantly worse than the $\mu\mu H$ channel, despite that the signal statistics in these two channels are larger. The $eeH$ channel has a significantly higher background than the $\mu\mu H$ channel, mainly from the single $W$ and single $Z$ backgrounds. Compared to the $\mu\mu H$ channel analysis, the event selection of the $eeH$ channel has a lower signal efficiency but three times larger SM background. As a result, the final accuracy of the $eeH$ channel is 80% worse than that of the $\mu\mu H$ channel.

In the $\nu\nu H$ channel, there is no direct measurement of the Z decay information, but severer backgrounds from the $ZZ$, $WW$, and $ZZ$or$WW$ processes. Therefore, the $\nu\nu H$ channel contributes the least accurate measurements among all these four channels.

The result is extrapolated to the ILC. The ILC will be operated at 250 GeV centre-of-mass energy with polarized beams; therefore, the signal and background cross sections are different from that at the CEPC. We calculated the cross sections and the expected number of events at the ILC. Assuming that the efficiency for each signal and background stays the same for the ILC and the CEPC, extrapolation results are shown in Table 9. Comparing with the result in ILC [14], the independent analysis in this paper leads to an accuracy 10% better, this improvement is mainly from the di-jet system information.

The precise measurement might be influenced by the systematic uncertainties caused by luminosity measurement, the fit procedure, and other experiment effects. There are several other experimental effects such as acceptance, uncertainties of the $\tau$ finding, jet energy corrections, or the influence of passive detector material. Further quantitative analysis of these effects is still needed. The uncertainty of the fitting procedure could be estimated by changing the background shape and

**Table 9** Extrapolated accuracy $\delta(\sigma \times BR)/(\sigma \times BR)$ in the ILC 250 GeV (2000 fb$^{-1}$)

|  | CEPC | ILC(L) | ILC(R) |
| --- | --- | --- | --- |
| Luminosity ($ab^{-1}$) | 5.6 | 2 | 2 |
| Polarization ($e^-, e^+$) | – | (0.8, −0.3) | (−0.8, 0.3) |
| Total Higgs | 1.18 M | 0.60 M | 0.40 M |
| Accuracy (%) | 0.8 | 1.09 | 1.21 |

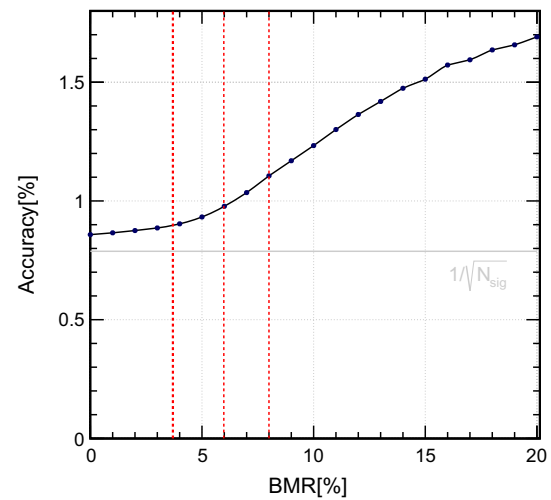

**Fig. 9** The dependence of the accuracy of the $qqH$ channel on the boson mass resolution. The thicker red line indicates that the CEPC baseline BMR is 3.8%, the other two red lines show the accuracy when the BMR drops to 6% and 8%. (Signal: $ZH \to qq\tau\tau$, Background: $ZZ \to qq\tau\tau$)

fitting range, and the difference in the measurement is taken as the systematic uncertainty.

The final accuracy of the $H \to \tau^+\tau^-$ signal strength measurement strongly relies on the precision of the invariant mass reconstruction of the jet system. The latter can be quantified using the Boson Mass resolution (BMR), which is defined as the invariant mass resolution of the $\nu\nu H$, $H \to gg$ events at 240 GeV centre-of-mass energy. The CEPC baseline detector, under the support of the Arbor reconstruction algorithm, essentially reaches a BMR of 3.8% [11,13]. A fast simulation is performed to quantify this dependence. This fast simulation takes into account the $qqH$ signal and the leading SM background ($ZZ$) after the event selection. It smears the 4-momentum of the identified Higgs decay final state, and predicts the accuracy of $qqH$, $H \to \tau^+\tau^-$ signal strength at different BMR, see Fig. 9.

For BMR better than 4%, the signal and background are well separated and the accuracy is only 10% worse than the statistic limit of $1/\sqrt{\epsilon_{sig} \cdot N_{sig}}$ (Here $\epsilon_{sig} \cdot N_{sig}$ refers to the number of signal events after the event selection, and $\epsilon_{sig}$ is the event selection efficiency). For a BMR between 4% to 15%, the accuracy degrades rapidly. For instance, the $qqH$ signal strength accuracy degrades by 20% if the boson mass resolution degrades from 3.8% to 8%. For BMR larger than 15%, the degrading tendency saturates, because the reconstructed invariant, and recoil mass of the Z boson does not provide significant separation power. In general, the BMR is essential for any measurement concerns the $qqH$ channel - the leading channel of Higgs signal at both CEPC and ILC, especially those with Higgs decays into non-jet final states. Therefore, a BMR better than 4% is highly appreciated for this benchmark measurement and should be pursued as an important goal for the detector design and optimization.

## 5 Conclusion

In this paper, the expected accuracies of the $H \to \tau\tau$ signal strength measurement of different channels with Higgs decaying into $\tau\tau$ at the CEPC and ILC have been studied. At CEPC, using full simulated samples, the combined accuracy reaches 0.8%. This result is extrapolated to the ILC, an accuracy of 1.2%/1.1% are anticipated for the ILC benchmark left/right polarization setting.

The signal events are divided into two kinds according to the number of jets. The $qqH$ channel dominates the accuracy. In the channels without jets, the accuracy of the $\mu\mu H$ channel is the best since the prompt $\mu$ pair provides clear signal signature, and is free of single Z background compared to the $eeH$ channel.

In the channels with a Z boson decaying to visible final states, the invariant mass of the $\tau$ pair using collinear approximation could efficiently distinguish the signal from the SM background. In the $\mu\mu H$ and the $eeH$ channel, fits on the invariant mass spectrum are used to determine the objective signal strength and improves the signal strength accuracy by 60%. In the $qqH$ channel, the cut on the invariant mass reduces the background by roughly 1 order of magnitude, at the cost of 30% of the signal efficiency.

A precise reconstruction of the impact parameter is essential for the $\tau$ events identification. In the $llH$ channel, the impact parameters are used in the TMVA training, and it can improve about 1/3 of the result compared to the TMVA without these parameters. In the $qqH$ channel, the fit on the impact parameter improves almost 50% with the final result of the cut chain.

The PFA oriented detector design and reconstruction are critical for this analysis. At channels without jets, including $\mu\mu H$, $eeH$, and $\nu\nu H$, the $\tau$ events identification relies strongly on a successful reconstruction of the photons and charged particles. The PFA oriented detector reconstructs the proper number of final state particles, providing critical multiplicity information. In the $qqH$ events, a dedicate $\tau$ finding algorithm is developed based on the precise reconstruction of final state particles. Equally importantly, the invariant and recoil mass of the identified di-jet system provides the separation between the signal and the backgrounds: the background could be suppressed by 1 order of magnitude, at the cost of 5% of the signal efficiency.

To conclude, the $H \to \tau^+\tau^-$ signal strength could be measured to accuracies of 0.8% and 1.1%–1.2% for the CEPC and ILC, respectively. Multiple analysis technologies and dedicated $\tau$ finding algorithms are developed in this manuscript. The PFA oriented detector design is critical for this measurement. A good precision on the mass reconstruction of hadronic final state is crucial for the analysis in the $qqH$ channel, and a BMR better than 4% is recommended, as one crucial performance benchmark for the detector design and optimization at those future $e^+e^-$ Higgs factories.

**Acknowledgements** The authors would like to thank Chengdong FU, Gang LI, and Xianghu ZHAO for providing the simulation tools and samples. This study was supported by the National Key Programme for S&T Research and Development (Grant NO.: 2016YFA0400400), the Hundred Talent programs of Chinese Academy of Science No. Y3515540U1. The work is supported by the Beijing Municipal Science & Technology Commission, project No. Z191100007219010. Y. Xu and X. Chen is supported by the National Natural Science Foudation of China (Grant NO.:11675087) and Tsinghua University Initiative Scientific Research Program.

**Data Availability Statement** This manuscript has no associated data or the data will not be deposited [Authors' comment: It is a simulation study with no experiment data].

## Appendix A: TAURUS(TAU reconstrUction toolS)

The package for $\tau$ finding in the CEPC is a double cone based algorithm, as shown in Fig. 10.

The steps for the $\tau$ candidates finding are: first, find tracks with energy higher than a defined $E_{min}$ as the seed; then collect tracks and photons within the small/larger cone (angle ConeA/ConeB) around the seed. The conditions for $\tau$ tagging are:

– an invariant mass of the particles in ConeA smaller than $M_{max}$,

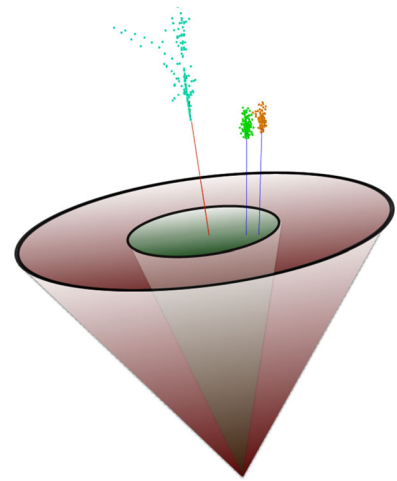

**Fig. 10** Double cone based $\tau$ finding algorithm in a $\tau \to \pi^-\pi^0$ event

– a number of tracks/photons in ConeA smaller than $NTrk$ and $NPh$,
– an energy ratio in the two cones greater than $R_{En}$.

Here the parameters $E_{\_min}$, ConeA, ConeB, $M_{max}$ and $R_{En}$ are optimized to the value $\epsilon \cdot p$, where $\epsilon$ is the efficiency of finding $\tau$ in qq$\tau\tau$ events (defined as the number of truth $\tau$ and found divided by the number of truth $\tau$), and $p$ is the purity of the tagged $\tau$s (defined as the number of truth $\tau$ and found divided by the number of tagged $\tau$). The value of these parameters are: $E_{min}$ = 1.5 GeV, ConeA = 0.15 rad, ConeB = 0.45 rad, $M_{max}$ = 2.0 GeV, $R_{En}$ = 0.92.

In this paper, after the $\tau$ candidates found, the invariant mass of the $\tau\tau$ system, the invariant mass of the qq system (the particles except for $\tau$s ) $M_{qq}$ and the recoil mass of the qq system (the particles except for $\tau$s ) $M_{qq}^{recoil}$ are used for the $H \to \tau\tau$ events.